\documentclass[pre,showpacs,preprint]{revtex4}

\usepackage{graphicx}
\usepackage{natbib}
\usepackage{amssymb}
\usepackage{amsmath}
\usepackage{verbatim}

\begin{document}

\title{Multiple nonergodic disordered states in Laponite suspensions:
a phase diagram }

\author{S. Jabbari-Farouji $^{1}$, Hajime Tanaka $^{3}$,
G. H.\ Wegdam $^{1}$ and Daniel Bonn$^{1,4}$  }

\affiliation{$^{1}$Van der Waals-Zeeman Institute, University of
Amsterdam, 1018XE Amsterdam, the Netherlands }

\affiliation{$^{2}$Theoretical Physics and Polymer Group, Department of Applied Physics, Technische Universiteit Eindhoven
 5600MB Eindhoven, The Netherlands}

\affiliation{$^{3}$Institute of Industrial Science, University of
Tokyo, Meguro-ku, Tokyo 153-8505, Japan }

\affiliation{$^{4}$ Laboratoire de Physique Statistique de l'ENS,
75231 Paris Cedex 05, France   }

\date{\today}

\begin{abstract}

We study the time evolution of different Laponite suspensions from
a low-viscosity ergodic state to a viscoelastic non-ergodic state
over a wide range of volume fractions and salt contents. We find
that the evolution of non-ergodicity parameter (Debye-Waller
factor) splits into two branches for all the samples, which
correspond to two distinct dynamically arrested states. At
moderately high salt concentrations, on the other hand, a third
and new nonergodic state appears that is different from the above
two nonergodic states. Measurement of the conductivity of Laponite
solutions in pure water shows that the contribution of counterions
in the ionic strength is considerable and their role should be
taken into account in interpretations of aging dynamics and the
phase diagram. Based on these data and available data in the
literature, we propose a new (non-equilibrium) phase diagram for
Laponite suspensions.
\\

{\it keywords:  Laponite, aging, colloidal glass, colloidal gel,
attractive glass}
\end{abstract}

\maketitle

\section{Introduction}

Understanding the phase behavior of clay suspensions is of
important technical and scientific interest. Characterization of
different ordered and disordered phases formed by clays is of
direct importance for various industrial applications such as soil
mechanics and for instance the control of viscoelastic properties
of materials with clay additives. Of more fundamental importance
is the study of the underlying mechanism behind gelation and glass
formation that are both observed in clays. This potentially
provides us with a deeper understanding of dynamically arrested
states of matter. Clay suspensions can typically be modelled as
charged anisotropic particles such as disks immersed in an
electrolyte, thus interacting via excluded volume, long-range
electrostatic repulsions and weak (van der Waals) attractions. The
phase diagram of anisotropic charged colloids such as clays and
understanding the aggregation, gelation and glass formation
appearing in such systems  is a matter of considerable debate
\cite{clay1,Mourchid,glass,Nicolai2005,Tanaka1,Tanaka2,PRL,Ruzicka2008}.
The specific clay system we study here, Laponite (a synthetic clay
\cite{Laponite}) has been the subject of intensive study over the
past decade or so.

Laponite consists of crystalline platelets with a thickness of 1
nm and an average diameter about 30 nm and a bulk density of 2.6
g/cm$^3$ \cite{Laponite}. Each Laponite particle is a three-layer
silicate composed of a central octahedrally coordinated
magnesium-oxygen-hydroxide sandwiched between two tetrahedrally
coordinated silica-oxygen sheets. Isomorphic substitutions of the
divalent magnesium atoms in the central layer by monovalent
Lithium atoms lead to the formation of negative charges within the
lattice, which is balanced by the sodium ions located at the
surface. When Laponite is dispersed in water or any
polar liquid, the polar molecules penetrate between interleaf
regions, dissolving the interleaf cations and separating platelet
surfaces by hydration and electrostatic forces. Thus in the final
suspension the Laponite surface has negative charge on the order a
few thousand electron charge (in water), while its edges (depending
on the pH) may have a small localized positive or negative charge
generated by desorption or absorption of hydroxyl group where the
crystal structure terminates.

For a range of Laponite concentrations and salt contents, the
dispersion of Laponite in water is followed by spontaneous
evolution from an
liquid-like state to a non-ergodic solid-like state. This process
is called aging, meaning that the physical observables of the
system such as diffusion of the particles and dispersion viscosity
evolve with time.

The aging dynamics during the fluid-solid transition in Laponite
suspensions has been independently studied by many groups now
\cite{Mourchid,willenbacher,kroon,Pignon,glass,Levitz,Bonn1,Bonn2,aging,Bellour,Nicolai2000,Nicolai2001,Nicolai,Nicolai2005,Cocard2000,Italian,Italian1,Sara, SaraMR,PRL,Harden,sarathesis,Joshi}
using light scattering and rheology techniques. Perhaps the
earliest studies on the phase behavior come from  Mourchid et al.
\cite{Mourchid} and Kroon et al. \cite{kroon}. Mourchid et al.
attempted to characterize the phase diagram of Laponite
suspensions based on rheological measurements \cite{Mourchid}.
They varied both particle concentration and ionic strength and by
performing oscillatory shear measurements on samples one week
after their preparation they defined a sol-gel transition line
where the zero frequency elastic shear modulus increases
remarkably. Their phase diagram gives a general overview, but we
have come to realize that their method is flawed as the
measurements were done after some arbitrary waiting time $t_w$ and
viscoelastic properties depend on $t_w$.

On the other hand, Kroon et al. \cite{kroon} studied the aging of
Laponite using dynamic light scattering (DLS) experiments. They
measured a range of sample concentrations between 2.2  and 3.5
wt\% and found that all the samples show a similar aging behavior
and evolve from an initially ergodic state to a non-ergodic state
around a certain time (ergodicity-breaking point detected by
changes in the moments of scattered intensity distribution) that
decreases exponentially with increasing concentration. For a  3
wt\% sample of Laponite, they reported the growth of the
non-ergodicity parameter (fraction of frozen-in density
fluctuations) from almost zero to approximately 0.8 at the late
stage of aging. Bonn et al. \cite{glass} suggested that aging in
samples with no added salt is due to strong electrostatic
repulsions, leading to the formation of a low volume fraction
Wigner glass. They determined the liquid-glass transition volume
fraction as a function of ionic strength, assuming that the
effective volume per particle can be estimated as $\pi R^2l_D$ (here $R\thickapprox 15 nm$ is the particle radius and $l_D$ is the Debye length)by
considering the Debye length  as the particle thickness. If this exceeds the
volume available per particle, which is estimated as $\pi R^2 h /
\phi$,($h=1$ nm the actual thickness $\phi$ is the volume fraction) no free volume is available and thus a glassy state may
emerge. The volume fraction for the liquid-glass transition is
given by $ \phi_{eff}\approx 0.5$ and thus varies as $\phi_{lg}
\sim h/l_D \propto I^{1/2}$ where $I$ is the ionic strength. Following the suggestion of Bonn et
al. \cite{Bonn2}, Levitz et al. \cite{Levitz}, by deionizing
Laponite indeed found the evidence for a Wigner glass at very low
ionic strengths consistent with the proposal of Bonn et al..

Nicolai and Cocard \cite{Nicolai2000,Nicolai} have studied the
aging at low concentrations of Laponite with added salt in the
ergodic regime of aging. They observed that the scattered
intensity increased with the aging time. At the late stages of
aging, when the scattered intensity has become constant, the
structure factor $S(q)$ shows a power-law q-dependence
characteristic of fractal structures for $S(q)$. In a later paper
the same group proposed a revised state diagram for Laponite
suspensions based on visual observations and waiting
time-dependent static light scattering experiments
\cite{Nicolai2005}. According to this study, the transition from
liquid to "solid" (defined as the state that does not flow when
tube containing the sample is tilted) occurs at much lower
concentrations than what was proposed in the phase diagram of
Mourchid et al. \cite{Mourchid}. From their observations they
argue that the origin of aging for all their measured samples with
salt and without salt ($C< 2 $ wt\%) is gelation rather than glass
formation \cite{Nicolai2005}.

The systematic study of Ruzicka et al.\cite{Italian,Italian1} was
the first one to show that non-ergodic states of Laponite can
exist at very low concentrations ($C\approx 0.3$ wt\%) of
Laponite, in the region which was proposed to be a sol according
to the phase diagram of Mourchid et al. \cite{Mourchid}. Using
DLS,  Ruzicka et al. \cite{Italian,Italian1} systematically
studied the aging dynamics of both low and high concentrations and
also varied the salt concentration. They showed that the intensity
correlation functions at low and high concentrations evolve in a
distinctly different manner and two distinct master curves have been identified. They suggested that aging at low
concentrations proceeds by formation of a Wigner glass made of
Laponite clusters, while at higher concentrations a glass is
formed whose basic unit is a single Laponite particle
\cite{Italian}. They show that this result is not affected by the
presence of salt and one still finds two distinct routes of aging
\cite{Italian1}. However, based on their recent small angle X-ray
scattering measurements,  Ruzicka et al. \cite{Ruzicka2008}
conclude rather that the arrested state at low concentrations
should be called a gel and at high concentrations an `attractive
glass'.

The study of Ruzicka et al. is the most comprehensive and
systematic one up to now. However, their experiments considered
only the ergodic regime of aging and they do not present any
results in the interesting range of concentrations $1.5 <C< 2.2$,
precisely the range between `dilute' and `concentrated' systems.
In a recent paper \cite{PRL}, for samples in the range $
0.1 <C < 3.6$  we showed that in the non-ergodic regime the
non-ergodicity parameter (Debye-Waller factor) also falls onto
either of two distinct master curves. This is consistent with the
classification obtained based on evolution of dynamic structure
factor in the ergodic regime of aging \cite{Italian}.The
evolution of non-ergodicity parameter provides us with some
valuable information about the nature of non-ergodic states. Using
this information altogether with other data such as short-time
diffusion and structure factor at low $q$ limit, we identified the
two observed distinct states as gels and glasses \cite{PRL}.
Furthermore we showed that for a range of intermediate colloid
concentrations $ 1.1 < C < 2.4 $, the transition to non-ergodicity
can occur in either direction (gel or glass). The distinction
between glass and gel was mainly made on the basis of (i) the
difference in the dynamic structure factor, notably the absence
(gel) or presence (glass) of cage-rattling motion; (ii) the
difference in static structure factor: $S(q)$ showed power-law
behavior as a function of q and a clear increase in time for the
gel state. Both are consistent with formation of some sort of structure (cluster or network)  in the gel,
and are absent in the glassy state.

It may be evident from these observations that the nature of
non-equilibrium phases formed by Laponite suspensions remains
ambiguous. Both gelation \cite{Mourchid,Nicolai2005} and glass
formation \cite{glass,aging,Italian,Joshi} have been proposed to account
for the aging process. Gelation and the glass transition have
important similarities. Both are ergodic to non-ergodic
transitions that are kinetic, rather than thermodynamic in origin,
and distinguishing between these two types of non-ergodic states
experimentally is a longstanding controversy
\cite{Tanaka1,Tanaka2,Mourchid,Nicolai2005,glass}. Here we will
show that at least part of the confusion about glassy or gel-like
behavior of Laponite suspensions  finds its origin in that each
group has only studied a specific range of concentrations or salt
content. Furthermore, most of the studies have been performed in
the ergodic regime of aging. Other serious discrepancies between
the results of different groups arise from the fact that some of
the measurements for determining phase diagram ignored the aging
features of Laponite and were done after some arbitrary waiting
time $t_w$ \cite{Mourchid,Levitz}. Another important confusion
regarding the phase diagram in the literature is the neglect of
the ionic strength that originates from the release of sodium ions
from the platelets when Laponite is dissolved in water \cite{glass}. The ionic
strength resulting from the release of sodium ions can be
estimated from conductivity measurements. We will show in the
following that the ionic strength of Laponite suspensions in pure
water is relatively large and cannot be ignored for determining
the `effective particle size' from adding the Debye length to the
size. Furthermore, different groups have used different grades
(XLG and RD) and batches of Laponite which also can change the
results slightly. Therefore this has lead to apparent
contradictions for the results reported by different groups, that
we attempt to clear up in the present paper.

Here we have extended our previous study \cite{PRL} to
Laponite samples with added salt, since an important part of the
discussion about the phase diagram is related to samples to which
salt is added to screen the electrostatic repulsion between the
Laponite particles. We report here extensive light scattering
measurements during the evolution from an initially ergodic
liquid-like state to a non-ergodic solid-like state on these
samples. We show that in the presence of moderate amounts of added
salt,in addition to the two distinct nonergodic states (A and B)
of Laponite, reported before \cite{Italian, PRL},  even a third
option (C) appears to exist for the system.

We discuss the nature of the three distinct dynamically arrested
states, A-C, and identify these, with the help also of data of
other groups in the existing literature, in different regions of
the phase diagram. This allows us to propose a unifying phase diagram for the non-equilibrium states of
Laponite.

\section{Experimental}

We study charged colloidal disks of Laponite XLG, with an average
radius of 15 nm and 1 nm thickness. Laponite can absorb water,
increasing its weight up to 20\%. Therefore, we first dried it in
an oven at $100^{o}$C for one week and subsequently stored it in a
desiccator.

We prepare a number of Laponite samples with different
concentrations and salt contents. Laponite solutions without added
salt are prepared in ultra pure Millipore water (18.2 M$\Omega$
cm$^{-1}$) and are stirred vigorously by a magnetic stirrer for one
hour and a half to make sure that the Laponite particles are fully
dispersed. The dispersions are filtered using Millipore Millex AA
0.8 $\mu$m filter units to obtain a reproducible initial state
\cite{glass}. This instant defines the zero of waiting time,
$t_{w}=0$.

The samples with added salt (NaCl, from Sigma) are prepared by
diluting the Laponite suspensions in pure water with a more
concentrated salt solution \cite{Nicolai2000}. For instance, a
sample of 0.8 wt \%, 6 mM NaCl is prepared by mixing equal volumes
of 1.6 wt\% Laponite solution in pure water with the same volume of
a 12mM salt solution.

A standard dynamic light scattering setup ($\lambda = 632.8$ nm) with a coherence factor close to 1 $(\approx 0.98)$
measures the time-averaged intensity correlation functions (Eq.
(\ref{eq:corr})) in VV mode, i.e. when polarization of incident
light and scattered light are both perpendicular (vertical)
relative to the scattering plane.

\begin{equation} \label{eq:corr}
g_{\mbox{\scriptsize t}}(q,t)
=\frac{<I(q,t)I(q,0)>_t}{<I(q,0)>_t^{2}}
\end{equation}
where $\langle \rangle_t$ stands for the time average. In the ergodic regime of aging
 this is related to the electric field correlation
function i.e., intermediate scattering function $f(q,t)$ through
the Siegert relation $g_{\mbox{\scriptsize t}}(q,t)=1+ \beta
|f(q,t)|^2$, where $\beta$ is an experimental factor close to
one \cite{Pecora}. For all the aging samples, there is a certain
 aging time after which the time-averaged correlation functions are
 no longer equal to their ensemble-averaged values, i.e. they change from one
position to another in the sample. This defines the
ergodicity-breaking point $t_{eb}$. This point is experimentally
determined as the waiting time for which the time-averaged
normalized second moment $<I(t)^2>_t/<I(t)>_t^2$ of the scattered
intensity $I(t)$ is not equal to one anymore, in other words the
measured normalized correlation function $g_{\mbox{\scriptsize
t}}$ does not decay from 1 to 0 anymore \cite{kroon}.

For aging
times $t_a > t_{eb}$, we calculate the ensemble-averaged electric
field correlation function i. e. intermediate scattering
function $f(q, t)$ from the time-averaged intensity correlation
function $g_t(q,t)$ and ensemble-averaged intensity
$I_E$ measured by rotating the sample at different heights
\cite{HS}.
\begin{equation} \label{eq:a1}
f(q,t)= 1+ (I_t/I_E)\{[g_t(q,t)-g_t(q,0)+1 ]^{1/2}-1 \}.
\end{equation}

The measurements are performed at scattering wave vector
$q=\frac{4\pi n}{\lambda}\sin (\frac{\theta}{2})$, where
$\theta=90^{o}$ is the scattering angle. The correlation functions
are measured at a rate depending on the speed of aging of
different Laponite suspensions.

\section{Results}

\begin{figure}[t]
\begin{center}
\includegraphics[scale=0.70]{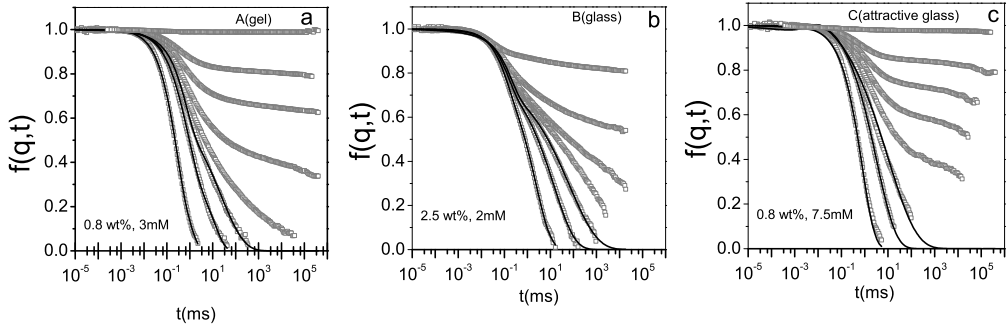}
\caption{Evolution of intermediate scattering function $f(q,t)$ for three
Laponite samples with salt at scattering angle of $90^{o}$. The
symbols present the measured correlation functions at increasing
waiting times (from left to right) that are ($t_{w}=$ 0.075, 5.7,
7.3, 8.8, 9.7, 11.9, 15, 19 and 500 days) for 0.8 wt\%, 3 mM, ($t_{w}=$11, 104, 153, 205, 255, 366 and 2854 min) for 2.5 wt\%, 2
mM salt and ($t_{w}=$9,44, 66, 90, 119, 164, 311 and 3900 min) for 0.8 wt\%, 7.5 mM salt. In all panels, the lines, on the curves that decay to
zero (ergodic stage), show the fits with
$f(q,t)=A\exp(-t/\tau_{1})+(1-A)\exp(-(t/\tau_{2})^{\beta})$.
 }\label{fig1}
\end{center}
\end{figure}

Measuring the intensity correlations of scattered light from a
large number of aging Laponite suspensions, one always observes
two regimes of aging in the evolution of the intensity correlation
functions. In the first regime the system is ergodic, whereas the
second regime corresponds to a non-ergodic (arrested) state. The
cross-over from the former to the latter is visible in the
experiments:  the time-averaged normalized correlation function no
longer varies between one and zero, i.e., a part of the degrees of
freedom are frozen in on the time scale of the measurement. The
waiting time for which the time-averaged correlation functions are
no longer equal to their ensemble-averaged values (i.e., their
values change from one position to another in the sample) defines
the \emph{ergodicity-breaking time} $t_{eb}$. Figure \ref{fig1}
shows the evolution of ensemble-averaged intermediate scattering
functions $f(q,t)$ for three different samples. In all the cases,
the correlation functions evolve from an ergodic state to a
non-ergodic state as the system ages. One can also observe that
this generic behavior is not affected by the presence of salt.

\begin{figure}[h!]
\begin{center}
\includegraphics[scale=0.60]{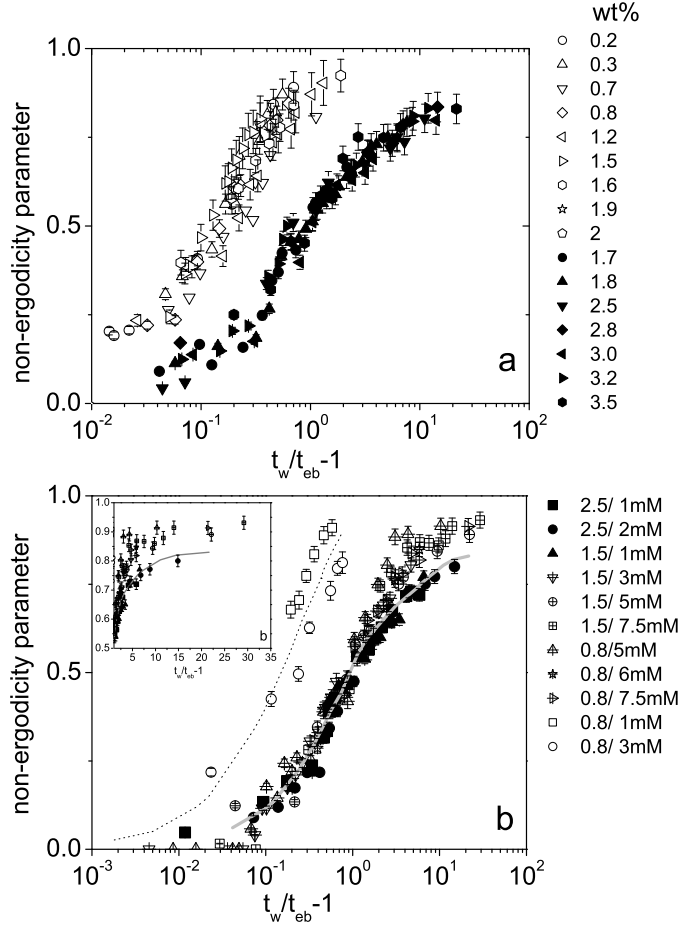}
\caption{The evolution of the non-ergodicity parameter
$f(q,\infty,t_w)$ versus reduced waiting time $t_w/t_{eb}-1$ for
different Laponite samples a) without salt b) with salt. The
colloid concentrations and salt contents are shown in the legend.
The samples can be divided into two groups according to the
evolution of non-ergodicity parameters. In part b the dashed and
solid lines correspond to the aging process of group A and B,
respectively, which are obtained from smoothed averaging over the
data of Laponite in pure water. As can be seen, for most of the
samples with salt (group C) the non-ergodicity parameter deviates
from the glass line at long waiting times. The inset of panel b
shows the difference in non-ergodicity parameter between the glass
(line, averaged over a large number of experiments) and the
attractive glass (data points). } \label{fig2}
\end{center}
\end{figure}

The intermediate scattering functions in the ergodic regime can be fitted with
the functional form
$f(q,t)=A\exp(-t/\tau_{1})+(1-A)\exp(-(t/\tau_{2})^{\beta})$, in which
$\tau_1$ and $\tau_2$ represent the fast and slow relaxation
times respectively \cite{Bonn1,Italian}. In the non-ergodic regime, the aging
rate of the system can be quantified by measuring the time
evolution of the non-ergodicity parameter $f(q,\infty,t_w)=\lim
_{t\rightarrow\infty} f(q,t_w)$ \cite{HS}. In the absence of salt we
have already shown that
 the evolution of non-ergodicity parameter in a range
of samples with different Laponite concentrations collapses onto
two distinct master curves when plotted as a function of reduced
waiting time $(t_{w}/ t_{eb}-1)$(Fig.\ 2a). These branches were
interpreted \cite{PRL} as belonging to a colloidal gel and
colloidal glass state based on  different aging behavior in other
measured quantities.
Here we refer to the branches for low and high concentrations
as A and B, respectively.
The observed differences between A and B
in the absence of salt can be summarized as follows.
\begin{itemize}
\item The static structure $S(q)$ of B changes very little with
waiting time, while that of A evolves dramatically. This is
due to formation of network-like structure or aggregation. This
difference manifests itself in the evolution of scattered
intensity with waiting time, which grows for A but is nearly
constant for B. \cite{PRL}\\

\item The short-time diffusion of particles in B decreases
only slightly while it drops significantly in A during the
ergodic to non-ergodic transition \cite{PRL} (See also Fig. \ref{fig4}a).

\item The slow relaxation time of B grows exponentially with
waiting time, while that of A grows faster than exponentially \cite{PRL,Italian}.

\item The distribution of relaxation times is different between A and
B: A has a broad distribution, whereas B has a double-peaked broad
distribution of relaxation times. \cite{PRL}.

\item The difference between A and B is perhaps best
reflected in their q-dependence of their structure factor at low
$q$ values and for late aging times. While the structure factor
of B is flat at low
q-values, $S(q)$ of A is q-dependent,
indicating that a structure has been formed \cite{PRL}.

\item The non-ergodicity parameter for A grows at a faster rate
than for B. While the non-ergodicity parameter for A
asymptotically reaches one, the non-ergodicity parameter for B
reaches an approximate value of 0.85 for late aging times,
indicating that there is still some freedom for the particles to
move. This is suggestive of the "cage rattling" picture of glassy
dynamics, also in agreement with the distributions of relaxation
times \cite{PRL}.
\end{itemize}

As explained in \cite{PRL}, in view of the above measurements, it
is tempting to identify A with a gel, and B with a glassy phase.
This will be discussed in detail below; we now first consider the
effect of added salt.

\begin{figure}[h!]
\begin{center}
\includegraphics[scale=1]{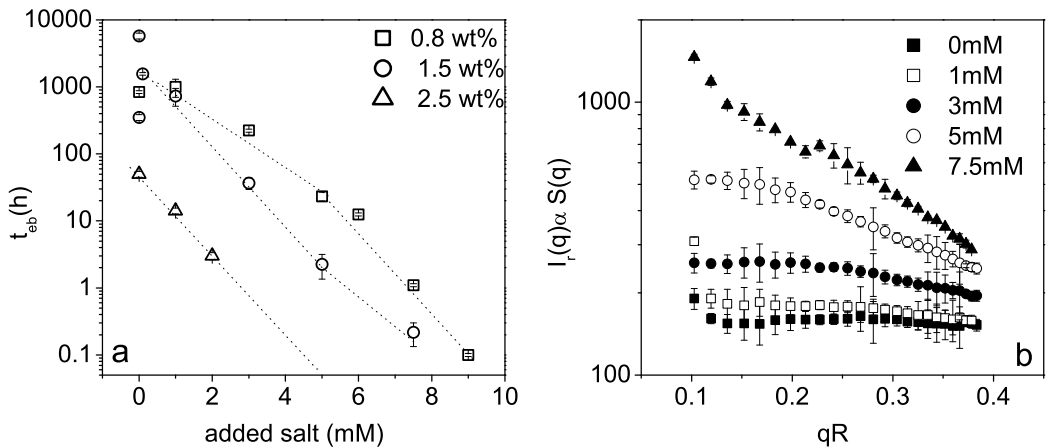}
\caption{(a) The ergodicity-breaking time $t_{eb}$ as a function
of the salt content for a few Laponite concentrations.  $t_{eb}$ decreases with adding salt. The dotted lines here are just guidelines for the eyes.
(b)The scattered intensity relative to the
toluene intensity $I_r$ as a function of dimensionless scattering
vector $qR$ for different amounts of added salt (NaCl) at a
concentration of 0.8 wt\%, as shown in the legends. These data are
taken a long time after the samples have become fully non-ergodic.
}
\label{fig3}
\end{center}
\end{figure}

It is clear that adding salt (NaCl) to a given concentration of
Laponite accelerates the aging. Figure \ref{fig3}a shows the
ergodicity-breaking time for the three samples as a function of
salt concentration. The effect is tremendous: by adding a few mM
of salt, $t_{eb}$ can decrease by 4 orders of magnitude, with a
roughly exponential dependence of the ergodicity breaking time on
 salt concentration.
We can see the change of the slope in Fig. 3(a)
for 0.8 wt\% around 5 mM salt concentration, which reflects the
crossover from A to C.

Furthermore, looking at the structure factor at the late stages of
aging (when the scattering intensity has been stabilized, $50 t_{eb}< t_w < 100 t_{eb}$) for different
salt contents, we find that with increasing salt the intensity
increases and the wave-vector dependence of $I(q)$ becomes more
pronounced (Fig. \ref{fig3}b). The observed change of the
structure factor with an increase in salt is plausible. The more
salt we add, the more we suppress the repulsive interactions,
therefore the attractive interactions play a more dominant role,
leading to formation of denser clusters and a more heterogenous
structure. This result is consistent with  the results of Refs.
\cite{Nicolai2001,Nicolai2005} in which the dependence of final
structure on salt content has been studied for several Laponite
concentrations (0.1, 1 and 1.5 wt\%).

Figure \ref{fig2}b shows that in spite of the accelerated aging, the
evolution of non-ergodicity parameter $f(q,\infty)$, versus scaled
waiting time $t_w/t_{eb}$ still splits into two branches similar
to those observed before for samples without salt \cite{PRL}.

However, looking carefully at Fig. \ref{fig2}b, it is evident that
some of the samples with salt deviate from the `glass' (B) master
curve (obtained from the data without salt) for longer waiting
times. These samples seem to evolve faster than the glass for long
waiting times $t_w > 3 t_{eb}$ and the non-ergodicity parameter
reaches values higher than for the glass, indicating a blocking of
rescaled particle motion. Measurements performed on these samples
at very long waiting times show that the non-ergodicity parameter
of these samples asymptotically reach the value 1. Hence, it turns
out that in the presence of salt, the story is more complicated
than the scenario sketched without salt \cite{PRL}.  If we look,
for instance, at the scattered intensity as a function of time
(Fig.\ \ref{fig4}), we find that for most of the samples with salt
the scattered intensity increases and concomitantly their
diffusion coefficient decreases with waiting time. Both increase
of intensity and decrease of short-time diffusion are in principle
indicative of the building up of structure, and thus suggest that
a gel forms \cite{PRL}: A=gel. Comparing, however, with the
`master curves' for the non-ergodicity parameter, we find that the
high-salt concentration samples (Lap 0.8 wt\% with 5 and 7.5 mM)
behave more like B, whereas the low-salt concentration samples
(Lap 0.8 wt\% with 1 and 3 mM) are on the master curve of A. All
of the 1.5 wt$\%$ Laponite samples should be glassy also,
according to the non-ergodicity parameter criterion; however at
least for the 1.5 wt\%, 7.5 mM sample also a clear increase in
intensity is observed.
\begin{figure}[t]
\begin{center}
\includegraphics[scale=0.75]{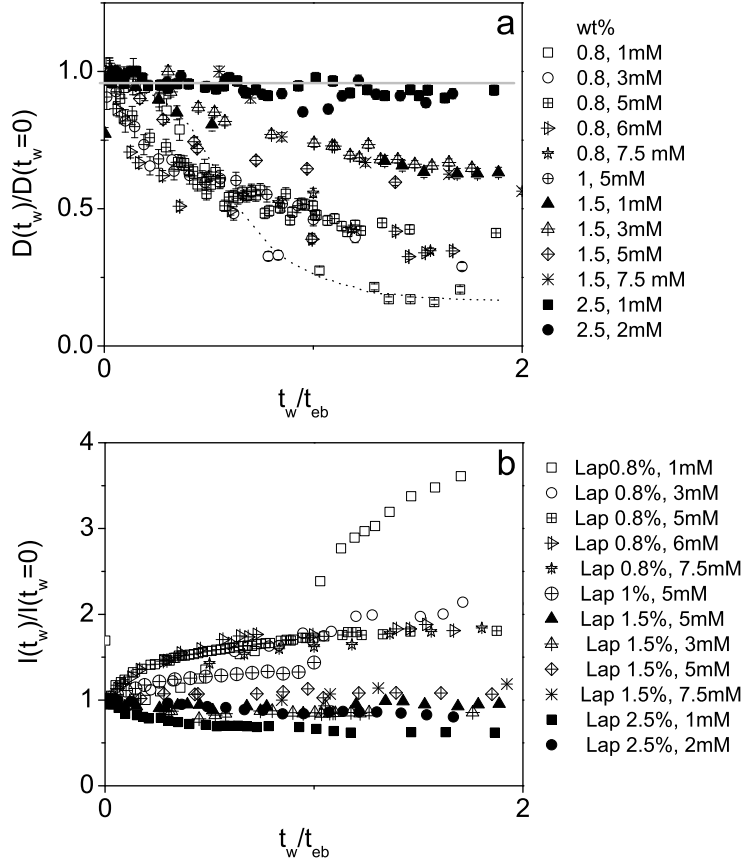}
\caption{(a) The evolution of short-time translational diffusion
normalized to its initial value $(t_w \approx 0)$ as a function of
$t_w / t_{eb}$. The solid and dashed lines show the B and A
line, respectively, obtained from smoothed averaging over the data
of Laponite in pure water. (b) Scattering intensity at scattering
angle $90^{o}$ as a function reduced waiting time. So as to focus
on the effect of aging, we have normalized the intensity to its
value at the beginning of aging.
}
\label{fig4}
\end{center}
\end{figure}

Plotting the slow relaxation
time $\tau_2$ normalized to its value at $t_w\approx 0$ as a
function of scaled waiting time $t_w/t_{eb}$, we find that
$\tau_2/ \tau_0$ for all the samples, with or without salt, splits
into two branches. We can see that for most of the samples
belonging to the B branch of non-ergodicity parameter the slow
relaxation time grows exponentially while for most of the samples
belonging to the A branch $\tau_2$ grows faster than
exponentially, similar to what observed for samples
without salt \cite{Italian,PRL}.
\begin{figure}[h!]
\begin{center}
\includegraphics[scale=0.8]{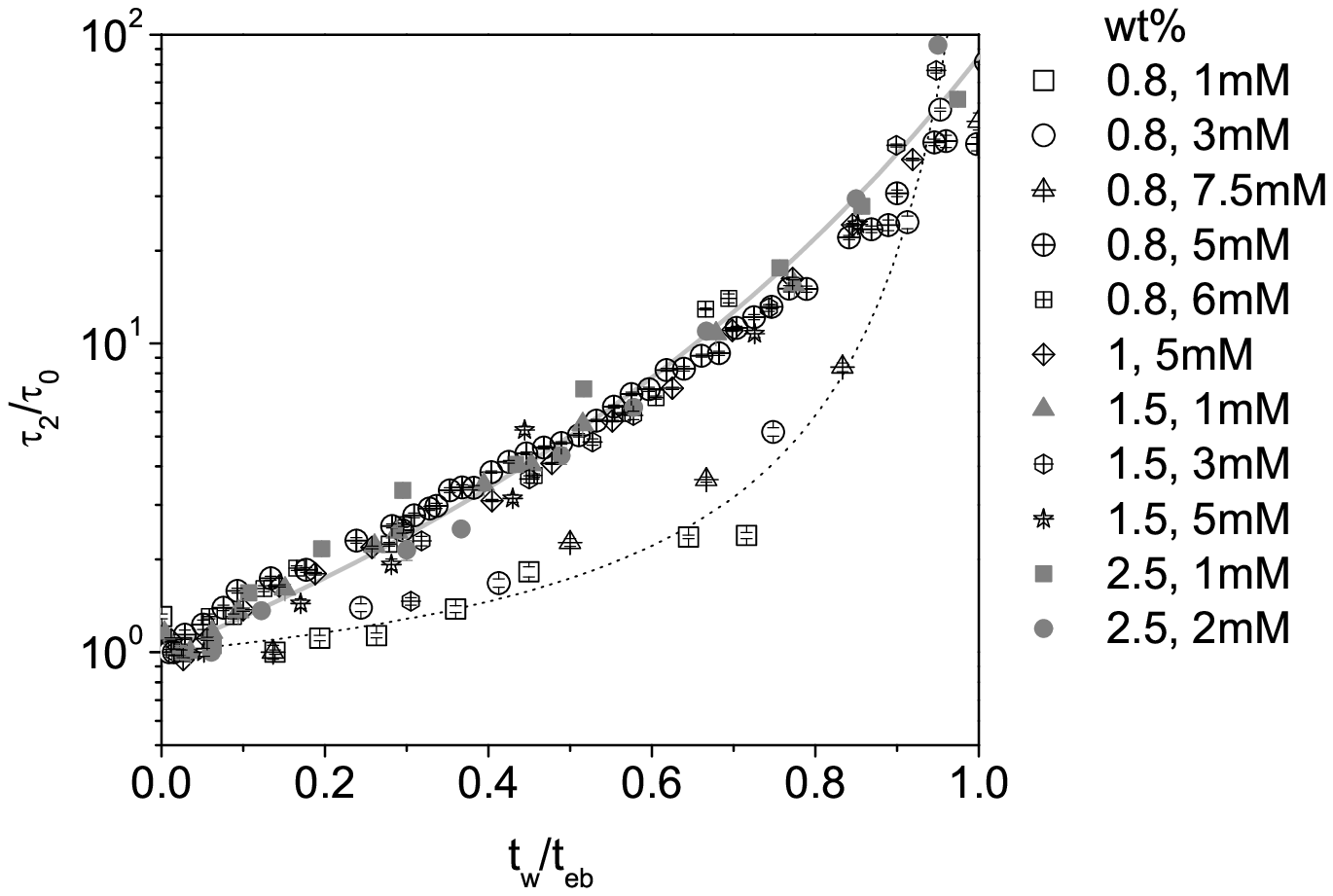}
\caption{The evolution of slow relaxation time as a function of
the scaled waiting time $t_w/t_{eb}$. The solid and  dashed lines
show the B and A line, respectively, obtained from smoothed
averaging over the data of Laponite in pure water.
}
\label{fig5}
\end{center}
\end{figure}

Comparing between different quantities, an inconsistency appears,
which is always the same one. Looking at the non-ergodicity
parameter, all the samples at moderate salt always behave like B.
However, some of their other measured quantities such as scattered
intensity, short-time diffusion and slow relaxation time,
consistently behave as if the sample were A. This situation is
indeed quite different from the one without salt as described in
\cite{PRL}. We see no `hesitations' of the samples between two
states in the sense that a sample that starts evolving in one
direction ends up in the other one. Rather, all individually
measured quantities consistently show an evolution in one
direction. The data therefore suggest that although the sample has
some definite characteristics of B, the other characteristics are
those of A.

Hence, to summarize, the addition of salt introduces new patterns
in the aging behavior in the sense that there are samples which
share some of the properties of A and some of the features of B.
We call this new set of samples C.  With this
classification, with added salt the samples Laponite 0.8 wt\% with
5 and 7.5 mM salt and Laponite 1.5 wt\%, with 3, 5 and 7.5 mM
belong to the new group C. The samples 2.5 wt\% with 1 and 2 mM
salt behave in all respects identically to the group B samples:
they are glassy. Also samples 0.8 wt\%, 1 and 3 mM behave exactly
like as group A samples, and therefore are gels.

\section{A phase diagram for non-equilibrium states of Laponite}

\begin{figure}[h!]
\begin{center}
\includegraphics[scale=1]{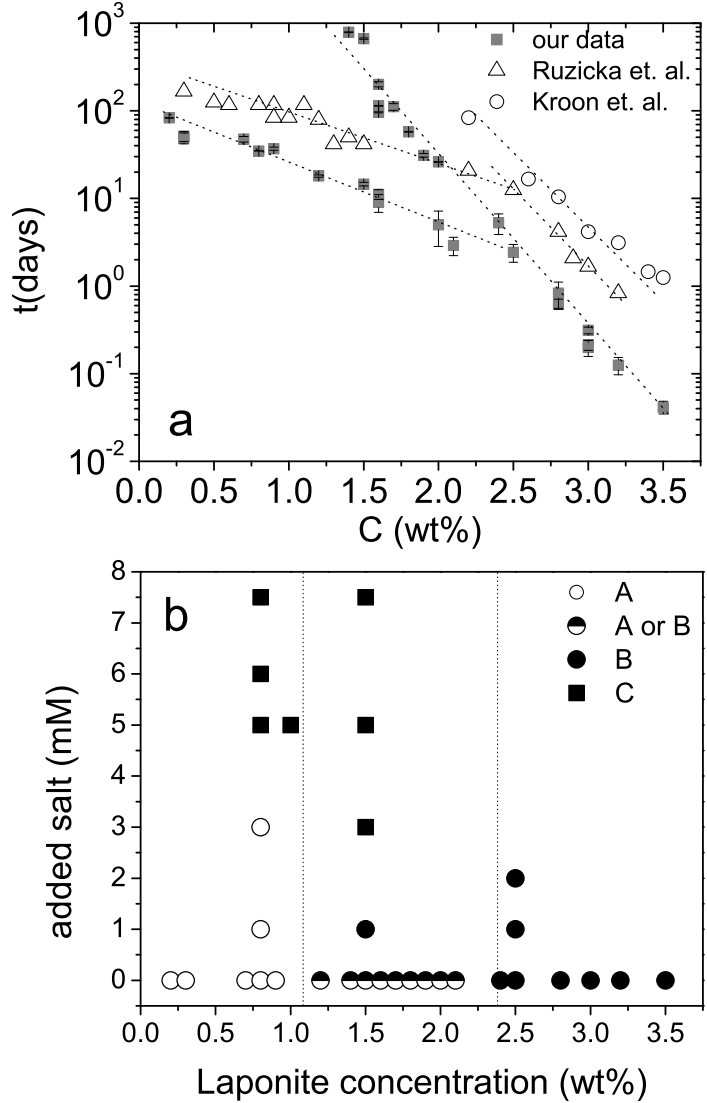}
\caption{(a) The ergodicity-breaking time $t_{eb}$ extracted from
our data and Kroon et al. data \cite{kroon} and
$t_w^{\infty}\propto t_{eb}$ extracted from Ruzicka et al. work
\cite{Italian} as a function of concentration of Laponite in pure
water. (b) Our proposed phase diagram based on light scattering
data for non-equilibrium states of Laponite.
}
\label{fig6}
\end{center}
\end{figure}

\begin{figure}[h!]
\begin{center}
\includegraphics[scale=1]{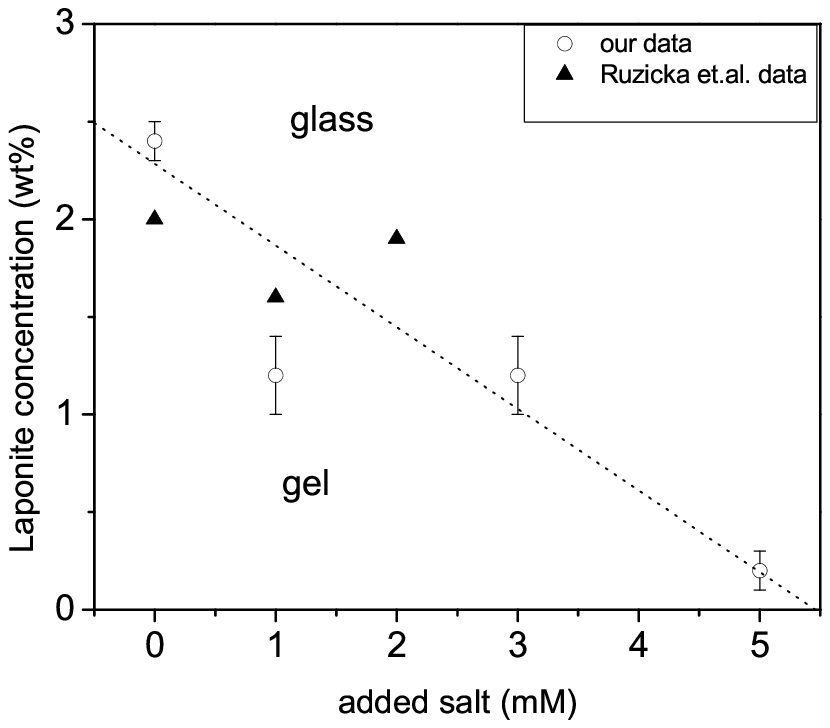}
\caption{The approximate A to B transition line obtained from our
data and data in Ref.  \cite{Italian1}. } \label{fig7}
\end{center}
\end{figure}

As we discussed in detail above, different groups have studied the
phase behavior of Laponite suspensions \cite{Mourchid,
Nicolai,kroon, Italian}, without a consensus emerging. Here, we
would like to show that one can get a consistent picture putting
all different pieces of information from different groups, despite
the fact that each group has used different batches of Laponite
and sometime different grades (Laponite XLG and RD). To
demonstrate this point, we have plotted in Fig. \ref{fig6}a the
ergodicity-breaking time obtained from our measurements
\cite{PRL}, from Kroon et al.'s experiments \cite{kroon} as well
as $t_w^{\infty}$ defined by Ruzicka et al. \cite{Italian}.
Ruzicka et al. obtained
 $t_w^{\infty}$ from fitting the waiting-time-dependent mean relaxation time $\tau_m$
with the general form $\tau_{m}(t_w)= \tau_0
\exp(B\frac{t_w}{t_w^{\infty}-t_w})$. Following the
proposal of Ruzicka et. al. \cite{Italian}, we fitted the mean
relaxation time from our data to the above form in order to
determine $t_w^{\infty}$. Comparing $t_w^{\infty}$ obtained from
our fits with $t_{eb}$ obtained directly from the experiments,  it
turned out that for lower concentrations $t_{eb}\approx
t_w^{\infty}$ and for higher concentrations $t_{eb}\approx 0.6
t_w^{\infty}$. Therefore, $t_w^\infty$ can also be interpreted as
a characteristic time for the transition from fluid-like to
solid-like state.
Figure \ref{fig6}a clearly shows that despite the difference in
aging speed which is due to different batches of Laponite, the
concentration dependence of ergodicity-breaking time found by the
different groups is very similar. The two main reasons for the
observed differences are most likely presence of additional water
in experiments of Kroon et al. \cite{kroon} (The Laponite was not
dried), and the difference in salt impurities between the very
pure XLG (our experiments) and RD (Kroon et al. \cite{kroon} and
Ruzicka et al. \cite{Italian}). A new phase diagram for
non-equilibrium states of Laponite based on our characterization
is shown in Fig. \ref{fig6}b. Our data suggest the existence of
three distinct nonergodic states A-C as demonstrated above. The
Ruzicka et al. data also reveal the existence of two different
arrested states (called IG1 and IG2 in their paper) for Laponite
samples in pure water and at low salt content. Indeed the Phase
IG1 of Ruzicka et al. corresponds to what we call A and their
phase IG2 to B. The consistency between their and our data becomes
even clearer if we plot the concentration at which the transition
from A to B occurs as a function of added salt, as depicted in
Fig. \ref{fig7}. As can be seen there is a fair agreement for
location of A to B transition line obtained from our data and the
Ruzicka et al. data.

\begin{figure}[h!]
\begin{center}
\includegraphics[scale=0.7]{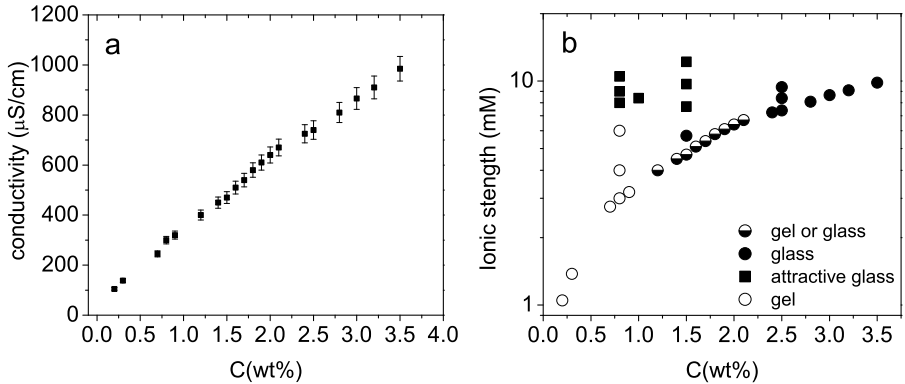}
\caption{(a) The conductivity of Laponite suspensions in pure
water as a function of concentration measured at early stages of aging, i.e $t_w \approx 0$. (b) The phase diagram with
modified salt axis for taking into account the ionic strength
resulting from counterions in the solution.
}
\label{fig8}
\end{center}
\end{figure}

Note that in the phase diagram of Fig. \ref{fig6}b the coordinate
($y$-axis) is the amount of added salt, while the interparticle
interactions between particles are controlled by the \emph{total}
number of  ions in the solution, i.e., the counterions released
from the surface of the Laponite particles plus the ions from the
added salt. To get an idea about the number density of counterions
from Laponite, we have measured the conductivity of Laponite
solutions. Figure \ref{fig8}a shows the conductivity of Laponite
solutions in pure water as a function of concentration
measured at early stages of aging. We also measured the
conductivity values for later stages of aging, before the samples
become solid-like. We observed only very small changes, at most an
increase in the conductivity of 5\% as a function of aging time
was found. The measured conductivity is mainly due to the
Na$^{+}$ counterions released from surface of Laponite particles.
The contribution of OH$^{-}$ ions released from the edges of
Laponite particles is relatively small. Neglecting this
contribution, the number density of Na$^{+}$ ions $n$ can be
obtained from $n_{Na}=\sigma_{Na}/\mu_{Na}e$, where $\mu_{Na}$ is
the mobility of $Na^{+}$ ions ($\mu_{Na}=5.19\times10^{-8 }$
m$^{2}$ s$^{-1}$ V$^{-1}$ \cite{CRC}) and $e$ is the electron
charge. The number density of Laponite particles can be estimated
as $n_{L}={\Phi_{\mbox{\scriptsize m}}}/{\rho_{\mbox{\scriptsize
L}}v_{\mbox{\scriptsize L}}}$ where $\Phi_{m}$ is the mass
fraction of Laponite particles, $\rho_{L}$ and $v_{L}$ are the
density and volume of an individual Laponite particle.
($\rho_{L}=$ 2.6 g/cm$^{3}$ is used here and $ v_{L}$ is
calculated assuming a disk radius of 15 nm and thickness of 1nm ).
This allows us to plot a new phase diagram (see Fig.\ \ref{fig8}b)
with $y$ axis being the total ionic strength $I=1/2\sum_i n_i
Z_i^2 $, in which $n_i$ is the number density of ion species $i$
and $Z_i$ is the charge of the ion species $i$. The abscissa is
the Laponite concentration. As can be seen the ionic strength
resulting from counterions is considerable and cannot be ignored.

Therefore the total ionic strength in Laponite solutions in pure
water is much higher than the 0.1 mM salt below which the Wigner
glass was predicted \cite{glass} and observed by Levitz et al.
\cite{Mourchid}. Taking into account the contribution of
counterions resolves the confusion about the absence or presence of Wigner
glass in Laponite suspensions in pure water. Levitz et al.
prepared Laponite solution of extremely low ionic strengths by
deionizing Laponite suspensions and immersing until the desired
ionic strength. As a
result they could observe a low volume fraction solid-like state.
Measuring the structure factor with ultra-small angle x-ray
scattering, they correctly identified this phase as a Wigner glass driven
mainly by long-range electrostatic repulsions.

\section{Discussion }

To summarize, we have shown that Laponite suspensions can form
different types of non-ergodic states (A-C)
upon changing concentration and salt content. We
have shown that the evolution of the non-ergodicity parameter
(Debye-Waller factor) falls into distinct branches for all
Laponite and salt concentrations.

Now we come to the most important question of what is the nature
of states A-C.
For samples without salt, there are the two branches, A and B.
The key difference is in the evolution of the
static structure factor and translational diffusion coefficient
with waiting time. The static structure factor and short-time
translational diffusion of B are independent of waiting
time, while the same quantities are strong functions of the waiting
time for A. In addition, the slow relaxation time in B
grows exponentially with waiting time, whereas it grows faster
than exponentially in A.

In group B samples (high Laponite concentrations, no or little added salt), the spatial structure is homogenous.
This group show similar aging patterns in the late stage
as seen in hard sphere glasses
\cite{HS} and simulations of Laponite in glassy state \cite{Mossa}.
For instance, even in the latest stages of aging particles maintain
their freedom of rattling in the cages formed by  their neighboring
particle, as evidenced by waiting time independent short-time diffusion
and a non-ergodicity parameter less than 1, which never exceeds 0.85 even
at the latest stage of aging.

\begin{figure}[h!]
\begin{center}
\includegraphics[scale=1]{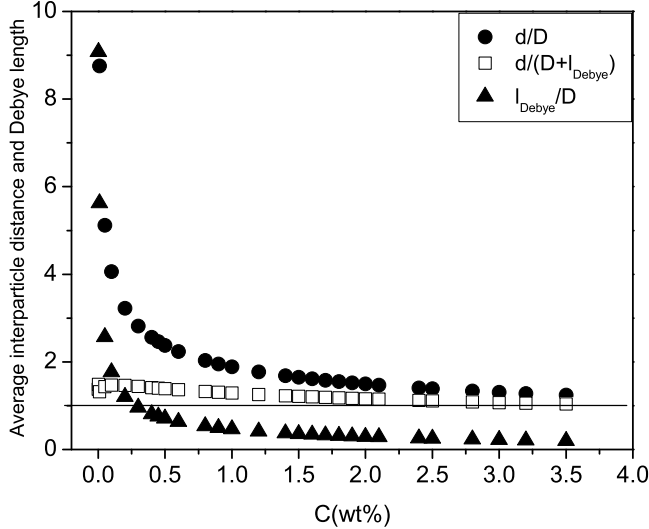}
\caption{ The ratio of average interparticle distance $d$ to particle
size $D=30$ nm and particle size plus Debye length $D+l_{Debye}$ as a
function of concentration for Laponite suspensions in pure water. }
\label{fig9}
\end{center}
\end{figure}

We suggest to call this group a repulsive glass in the same sense as
the glass formed at high concentrations of hard spheres. Note that
although both attractive and repulsive interactions are present in all
ranges of Laponite and salt concentrations, we believe that attractions
do not play a dominant role at these relatively high concentrations, as is evidenced by
the homogenous structure of these suspensions. This analogy becomes
clearer if we plot the ratio of average interparticle distance between
Laponite particles $d$ to particle size $D=2R=30$ nm versus
concentration (See Fig. \ref{fig9}). $d$ is estimated as $(\pi R^2 h / \phi)^{1/3}$,
where $\phi$ is the volume fraction of Laponite particles.
Figure \ref{fig9} clearly demonstrates that $d$ is very much comparable
to particle diameter $D$. Adding the Debye length to the particle size
makes this correspondence even better. Our interpretation of this glassy
state at high concentrations is a jammed state which is appearing
at much lower volume fractions compared to spheres due to anisotropic
shape of disks and their large excluded volume effects. Interestingly
Fig. \ref{fig9} correctly pinpoints the onset of deviations from
glassy behavior ($C\approx 2$ wt\%).

 Recently Ruzicka et al.
\cite{Ruzicka2008} assigned group B as attractive glass.
With this assignment,
the transition from gel to glass can be explained simply by that
the increase in the volume fraction of the particles leads to the
decrease in the void size in gel and eventually the void size
decreases nearly to the particle size (attractive glass).
Then, however, this scenario can explain neither the absence
of the slowing down of the single-particle diffusion for group B
nor the absence of the increase in the scattering intensity
since attractive glass should accompany
the finite-time bond formation between particles and
slight increase in the scattering intensity.
From this consideration, we suggest that the scenario that
group B is a repulsive glass is more plausible.

The assignments of groups A and C are much more subtle. In group A
samples (low concentrations, no or little added salt),
the aging behavior is distinctly different from that of group B. A
dramatic decrease of short-time diffusion and remarkable increase
of scattered intensity with waiting time is observed. The final
structure in such samples is very heterogenous, suggesting the
formation of a structure in the form of cluster or gel network.
These samples are also characterized by a non-ergodicity parameter
which reaches the value 1 roughly at a time  equal to twice the
ergodicity breaking time. All these features are consistent with the
formation of a gel of Laponite particles, due to attractions
between the particles \cite{PRL}. The assignment of group A as gel
has been made by Jabbari-Farouji et al. \cite{PRL} and more recently also by
Ruzicka et al. \cite{Ruzicka2008} As shown in Fig. 9, it is
obvious that in this range the repulsive interactions are not
enough to stabilize a repulsive glass. Nevertheless, it is not so
clear why a gel formed by attractive interactions appears in the
region of weaker screening of repulsive interactions than the
glass. The anisotropic shape of Laponite together with its rim of
opposite charges (or neutral) may result in the aggregation in the
concentration range. The formation of gel in such a very dilute regime may be
a consequence of the competing attractive
and repulsive interactions \cite{Coniglio,gel1}.
Thus, we conclude that A is indeed a gel.

There is a third group of samples, group C samples (moderate
Laponite concentrations and high salt content), which also show a
heterogenous structure but their aging behavior shares some of the
features of group B (glass) and some of the group A (gel). For
example their scattered intensity increases with waiting time (a
characteristic of a gel) while their non-ergodicity parameter
evolution shows a similar behavior to that of group B (a glass).
However unlike group B, the non-ergodicity parameter of the group
C does not saturate at a lower value than 1, but keeps on
increasing to reach the value 1 asymptotically at very large $t_w
> 10 t_{eb}$ .


Thus, samples in group C share some of the features of the glass, and some of
the gel. This could be due to the fact that particles aggregate to
form clusters, the diffusion of which becomes hindered
progressively as the clusters grow. It has indeed been
proposed that such a `cluster glass'  exists
\cite{clay1,Italian,weitz,Sciortino,Kroy}, for which the size of
clusters grows in time; this in turn makes that their diffusion
significantly slows down. Together with the small amplitude of the
motion, the diffusion mode inside a cage may become more and more
difficult to observe. This may explain why the non-ergodicity
parameter reaches 1 at very late waiting times. In suspensions of
a similar type of clay (monmorinite), Schurtenberger and his
coworkers \cite{clay1} found `cluster fluids' in the corresponding
dilute region of the phase diagram.

Our experiments show that particles diffusion is suppressed
in both group A and
C at late times. However, we may need to take special care when
interpreting the information of Fig. 4. When we add salt, it
is expected that even at $t_w=0$ (in our definition) some clusters
may have already formed. For example, the scattering intensity at
90$^\circ$ increases more steeply for group A than for group C,
but the final scattering intensity is lower for group A than for C
(see Fig. 3b). This means that at $t_w=0$ the samples are more
heterogeneous for group C than for A. This may be natural on
noting that the attractive interaction inducing aggregation is
stronger for group C than for A. On the other hand, C has characters of both glass and gel, where
the slowing down of the dynamics is both due to aggregation (as in
a gel) and steric hindrance (as in a glass).
The coexistence of these two characters may be quite naturally explained
by phase-separation-induced dynamic arrest:
phase separation leads to a dense region of Laponite,
whose composition is high enough for the formation of attractive glass.
Such a scenario was proposed
for suspensions of uncharged colloids \cite{Manley}.
The strong heterogeneity of the state C is also compatible with
this scenario.
So we assign the state C as attractive glass formed
by phase separation (arrested phase separation due to the
formation of attractive glass).

 Phase separation requires the presence of
attractive interactions. At this moment the nature of the
attractive interactions between Laponite particles is unclear.
Possible sources are the van der Waals interactions and the
attractions between the positive charge on the rim and the
negative charge on the surface of Laponite particles. Indeed,
recent experiments \cite{lili} have shown evidence for a
short-range attractive potential in the effective interaction
energy.

At this point it is worthwhile comparing the aging features of our
`attractive glass' induced by phase separation
with other attraction-driven glassy systems such
as the attractive glass formed in hard spheres with added short-ranged
attractions \cite{pham,AttExp2,cates}. In hard sphere systems with
added attractions, attractive glasses are formed
at moderately high volume fractions of particles and for strong
enough attractions. This can be achieved in experiments by adding
polymers that cause a depletion interaction \cite{pham,AttExp2}.
Light scattering studies in these systems have revealed significant
differences between
attractive and repulsive glasses in both their static and dynamic
properties. Pham et al. \cite{pham} showed that upon increasing
the attraction strength, entering the attractive glass region
(for a fixed volume fraction of colloids), the peak position of
the structure factor shifts to a higher $q$-value and its height
slightly decreases.  This shows that average interparticle
distance is decreased upon increasing the attractions and
particles bond in clusters, implying that the average number of nearest
neighbors should decrease (leading to the decrease in peak
height), and `holes' open up which render the structure more
heterogenous on a length scale of a few particles. The
increased heterogeneity is reflected in a slight
increase of the structure factor at low $q$ values. A similar
trend is observed in our data upon increasing the salt
concentration that screens the repulsions; this should be equivalent to
increasing the attractions in the colloid-polymer system.

The differences in dynamics of attractive and repulsive glasses is
most evident in the short-time relaxations
\cite{pham,cates}. The short-time dynamics of particles
progressively departs from free diffusion upon increasing the
attraction. In fact, for attractive glasses the particles are
confined so tightly by attractive potential wells that short-time
diffusion drops dramatically compared to the repulsive glass at
the same particle concentration \cite{pham}. This is consistent
with our attractive glass samples for which a decrease of the
short-time diffusion is observed (Fig. \ref{fig4}).
Thus, our attractive glass shares some important similarities
with attractive glasses in other systems.

\section{Conclusion}

To summarize, we report that there are at least three
distinct types of dynamically arrested states in Laponite
suspensions. We specify the aging process towards these non-ergodic
states in detail, using both static and dynamic light scattering.
Our data indicate that the competition between short-range (van
der Waals) attractions and long-range (electrostatic) repulsions
in anisotropic Laponite particles leads to a rich non-ergodic state
diagram and corresponding aging behavior. On the basis of our data
on the static structure factor and the dynamics of the aging, in
conjunction with other observations in the literature, we propose
that the three observed distinctly different arrested states
should be identified as gel (A), repulsive glass (B) and
attractive glass (C).

The gel state is formed at low clay concentrations and low amounts
of added salt. It has a spatially heterogenous structure as
evidenced by our static light scattering measurements. In this
case the aggregation of particles either in the form of a
network-like structure or clusters is responsible for the aging
process. The main characteristics of aging in a gel are dramatic
slowing down of translational diffusion and a fast growth of
non-ergodicity parameter to a fully non-ergodic state specified
by non-ergodicity parameter of value 1.

The glassy state forms in relatively high concentrations of
Laponite and low amounts of added salt. Here, the aging dynamics
of a glass has its origin in cage-diffusion process: for short
times or small displacements 'normal' Brownian motion is observed;
however, for larger times or excursions, the particles are
confined in effective cages formed by their neighbors. This
becomes more and more difficult as time goes on, due to the fact
that the system finds deeper and deeper energy minima during the
aging process. On the other hand, even for late times the
particles maintain their free rattling in the cage, as evidenced
by a waiting-time independent short-time diffusion and
non-ergodicity parameter different from 1 even at the latest
stages of aging.

Our study also suggests that a third non-ergodic state exists in
Laponite suspensions, which we call attractive glass. This state
is formed for moderately high amounts of salt shares some features
of a glass and some of the gel. It has a heterogenous spatial
structure similar to a gel while its dynamics is something between
that of a gel and a glass. 

It is interesting at this point to discuss the relation
of our light scattering measurements on these non-ergodic states
with their rheological properties. Most interesting of course is
to see whether a difference between the two 'glassy' states:
attractive (B) and repulsive (C) glass can be found. As described
in detail in Ref. \cite{SaraMR}, we performed local microrheology
(MR) experiments on samples belonging to groups B and C. In this
technique, one looks at the (Brownian) motion of a probe particle,
from which the visco-elastic properties of the surrounding medium
can be inferred. It was found that that although the complex shear
modulus shows a very similar frequency dependence for both types
of samples, the local MR measurements reveal the differences in
the structure. Local shear moduli obtained from samples of group B
are independent of position in the sample while for a sample in
group C, a significant heterogeneity in the sample develops as the
sample ages. Therefore the shear moduli differ from one position
to another in the sample. This provides one more piece of evidence
for the classification proposed here.

\section*{Acknowledgments}
The research has been supported
by the Foundation for Fundamental Research on Matter (FOM), which
is financially supported by Netherlands Organization for
Scientific Research (NWO). LPS de l'ENS is UMR8550 of the CNRS,
associated with the universities Paris 6 and 7.
H.T acknowledges a partial support from
a grant-in-aid from the Ministry
of Education, Culture, Sports, Science and Technology, Japan.

\end {document}